\documentclass[notitlepage,nofootinbib,superscriptaddress,showpacs,aps,prd]{revtex4-1}

\usepackage{amsmath}
\usepackage{amsfonts}
\usepackage{amsthm}
\usepackage{amssymb}
\usepackage{graphicx}
\usepackage{epsfig}
\usepackage{hyperref}
\usepackage{amscd}

\usepackage[T1]{fontenc}
\usepackage{ae,aecompl}
\usepackage{amssymb}
\usepackage{multirow}
\usepackage{booktabs}
\usepackage{chngpage}
\usepackage{float}
\usepackage{setspace}
\usepackage{graphicx}
\usepackage{amsmath}	
\usepackage{amssymb}	

\begin{document}
\title{Test  particles in a magnetized conformastatic   spacetime}

\author{Antonio C. Guti\'errez-Pi\~{n}eres}
\email{gutierrezpac@gmail.com}
\affiliation{Facultad de Ciencias B\'asicas, Universidad Tecnol\'ogica de Bol\'ivar, Cartagena, CP 131001, Colombia}
\affiliation{Instituto de Ciencias Nucleares, Universidad Nacional Aut\'onoma de M\'exico,  AP 70543,  M\'exico, DF 04510, M\'exico}

\author{Abra\~{a}o J. S. Capistrano}
\email{abraao.capistrano@unila.edu.br}
\affiliation{Federal University of Latin-American Integration, 85867-670, P.o.b: 2123, Foz do iguassu-PR, Brazil}
\affiliation{Casimiro Montenegro Filho Astronomy Center, Itaipu Technological Park, 85867-900, Foz do Iguassu-PR, Brazil}

\author{Hernando Quevedo}
\email{quevedo@nucleares.unam.mx}
\affiliation{Instituto de Ciencias Nucleares, Universidad Nacional Aut\'onoma de M\'exico,  AP 70543,  M\'exico, DF 04510, M\'exico}
\affiliation{Dipartimento di Fisica and ICRANet, Universit\`a di Roma ``La Sapienza"  I-00185 Roma, Italy}

\begin{abstract}
A class of   exact   conformastatic  solutions of  the Einstein-Maxwell field equations is presented in which the gravitational and electromagnetic
potentials are completely determined by a harmonic function.  We derive the  equations of motion for  neutral and charged  particles in a spacetime
background characterized  by  this  class of solutions.  As an example, we focus on the analysis of a particular harmonic function, which generates
a singularity-free and asymptotically flat spacetime that describes the  gravitational field of a punctual mass endowed with a magnetic field.
In this particular case, we investigate the main physical properties of equatorial circular orbits. We show that due to the electromagnetic interaction, it is
possible to have charged test particles which stay at rest with respect to a static observer located at infinity.
Additionally, we obtain an analytic expression for the  perihelion advance of test particles, and the corresponding explicit value in the case of a punctual
magnetic mass. We show  that the analytical expressions obtained from our analysis are sufficient for being confronted with observations, in order 
to establish whether such objects can exist in Nature.

\end{abstract}


\pacs{04.20.Jb, 04.40.Nr}

\maketitle


\section{Introduction}\label{sec:Introduction}
In recent years, the interest in studying  magnetic fields has increased in both astrophysics and cosmology. In astrophysical dynamics, the study of disk sources for stationary axially symmetric spacetimes with magnetic fields is of special relevance mainly on the investigation of neutron stars, white dwarfs, and galaxy formation. In this context, usually, it is
believed that electric fields do not have a clear astrophysical importance; nevertheless, there is a possibility that some galaxies are positively charged \cite{gonzalez2008finite, bally1978electrically}. On the other hand,
magnetic fields are very common in astrophysical objects, and can drastically  affect other physical properties (e.g, H-alpha emission, density mass, local shocks, etc.). For instance, magnetic fields
in a galaxy can be measured from the non-thermal radio emission under the assumption of equipartition between the energies of the magnetic field and the relativistic particles (the so-called energy
equipartition); this interaction can play an important role on the formation of arms in spiral galaxies \citep{uyaniker2004magnetized}. For nearby galaxies, one can use other effects such as  optical polarization,
polarized emission of clouds and dust grains, maser emissions, diffuse radio polarized emission and rotation measures of background polarization sources as well.  In the case of the Milky Way, e.g.,
the magnetic field has been actively studied in its three main regions (central bulge, halo and  disk). Moreover, magnetic fields seem to play an important role on the formation of jets
(resulting from collimated bipolar out flows of relativistic particles) and accretion disks near black holes and neutron stars
\citep{zamaninasab2014dynamically}.

It is important to stress that  magnetic fields are found mainly on interstellar medium, remarkably, in spiral galaxies \citep{han2012magnetic} which can be described with a good approximation by means of thin
disks. The magnetic and gravitational field of such objects can reach very high values. In a series of recent works
\citep{GRG2015ACG-P2015,  gutierrez2015exact, gutierrez2015variational}, several classes of static and stationary axisymmetric exact solutions of the Einstein-Maxwell equations were derived, which can be
interpreted as describing the gravitational and magnetic fields of static and rotating thin disks. Although these solutions satisfy the main theoretical conditions to be considered as physically
meaningful, additional tests are necessary in order to establish their applicability in realistic scenarios. For instance, the study of the motion of test particles and the comparison of the resulting
 theoretical predictions with observations are essential to understand the physical properties of the solutions and the free parameters entering them. This is the main goal of the present work.

To investigate the gravitational field of isolated non-spherical bodies, axial symmetry is usually assumed. This is a very reasonable and physically meaningful assumption. Nevertheless, it is clear
that Nature does not choose a particular symmetry, but, instead, we assume symmetries in order to simplify the mathematical
difficulties that are normally found when trying to find exact solutions for the description of gravitational fields. To investigate the possibility of having conformally symmetric compact objects in Nature, we explore in this work one of the simplest conformsymmetric solutions for an isolated body endowed with a magnetic
field.
We will see, in fact, that the resulting effects are enough to determine whether an isolated body belongs to the general class of axisymmetric compact objects or to its subclass of conformsymmetric objects.

In this work, we  follow the original terminology introduced by Synge \citep{synge1960relativity}, according to which conformastationary spacetimes are stationary spacetimes with a conformally flat
space of orbits, and conformastatic spacetimes comprises the static subset.  In a previous work  \citep{gutierrez2015motion}, a static conformastatic solution of Einstein-Maxwell equations was
presented and, in particular, the corresponding geodesic equations were derived explicitly. In the present work, we perform a detailed analysis of the equations of motion of test particles moving in a
conformastatic spacetime, which describes the gravitational and magnetic fields of a punctual source. In particular, we analyze the physical properties of circular orbits on the equatorial plane of the
gravitational source.  Additionally,  we find   an  expression  for   the  perihelion advance  in  a  general   magnetized conformastatic  spacetime.

This  work  is  inspired  by the approach presented in   references \citep{PhysRevD.83.104052} and
\citep{PhysRevD.83.024021}. The  analysis  presented  here serves as a ``proof-of-principle"  that gives a solid footing for a fuller study of particles motion  in the field of relativistic disks and for a later study in  even more realistic astrophysical context.

This  work is  organized  as  follows.  In  Section \ref{sec:Basic framework},  we present the general line element for conformastatic
gravitational fields which is a particular case of the general axisymmetric static line element. The  Einstein-Maxwell  equations are calculated explicitly and we show that there exists a class of solutions generated by harmonic functions. Moreover,
we derive the complete set of differential equations and first integrals that govern the dynamics of charged test particles moving in a conformastatic spacetime.
In addition, we show that, due to the spacetime symmetries, the geodesic equations on the equatorial plane can be reduced to one single ordinary
differential equation, describing the motion of a particle  in an effective potential, which depends on the radial coordinate only.  In  Section \ref{sec:The  circular  orbits},  we derive the explicit
expressions for the energy and angular momentum of a particle moving along a circular orbit.   In  Section \ref{The  circular motion   in the  field  of  a  punctual mass in Einstein-Maxwell  gravity}, we  focus on the particular case of a punctual source.
We analyze  all the physical and stability  properties of circular orbits along which charged test particles are moving.
In Section \ref{sec:The perihelion advance in a  conformastatic magnetized  spacetime}, we obtain an expression for the  perihelion advance of a charged test
particle  in a generic  conformastatic spacetime  in the presence of  a magnetic field.
In this  section, we  also  consider the particular case of a punctual source to obtain concrete results that are confronted with the ones obtained in Einstein gravity alone.
Finally,  in  Section \ref{sec:conclusions},  we  present the  conclusions.

\section{Basic framework}\label{sec:Basic framework}
In Einstein-Maxwell gravity  theory,  a particular class of a gravitational fields can  be  described  by the  conformastatic metric in cylindrical coordinates
\citep{PhysRevD.87.044010}
\begin{equation}
      ds^2 = - c^2 e^{2\phi} dt^2 + e^{-2\phi} (dr^2 + dz^2 + r^2d\varphi^2), \label{eq:metric}
\end{equation}
where  $c$ is the  speed of   light in  vacuum   and the  metric  potential $\phi$ depends only on the variables $r$  and  $z$.
This represents a subclass of the axisymmetric static Weyl class of gravitational fields.
The  field equations are  the  Einstein-Maxwell equations
\begin{equation}
     R_{\alpha\beta}
                    - \frac{1}{2}g_{\alpha\beta}R   = k_{_0} E_{\alpha\beta},\quad
     \nabla_{\beta}F^{\alpha\beta}=0,
     \label{eq:E-Mequations}
\end{equation}
where  $k_{_0}= 8\pi\ $G c$^{-4}$
\footnote{Along this  work  we  use  CGS units such that  $k_{_0}= 8\pi\, G c^{-4} =2,07\times10^{-48}$
s $^2$cm$^{-1}$g$^{-1}$,  $G= 6.674\times10^{-8}$cm$^{3}$g$^{-1}$s$^{-2}$
and  $c=2.998\times 10^{10}$\,cm\,s$^{-1}$. In plots, however, for clarity we use geometric units with $G=c=1$.  
Greek indices run from 1 to 4.}.
The  energy-momentum  tensor $E_{\alpha\beta}$ is  given  by
\begin{equation*}
    E_{\alpha\beta} = \frac{1}{4 \pi} \left\{ F_{\alpha\gamma}F_{\beta}^{ \;\; \gamma}- \frac{1}{4}g_{\alpha\beta}F_{\gamma\delta}F^{\gamma\delta} \right\}, \label{eq:tab}
\end{equation*}
where the electromagnetic  tensor  is denoted by $F_{\alpha\beta} = A_{\beta,\alpha} - A_{\alpha\beta}$, being  $A_{\alpha} =(A_{t}, \mathbf{A})$ the electromagnetic four-potential. The  components
of the  electromagnetic four-potential depend  on $r$ and $z$ only.

For the sake of simplicity, we  suppose that  the only  nonzero component  of  the
four-potential  is $A_{\varphi}$. From a physical point of view, this means that we will limit ourselves to the analysis of
magnetic fields only; electric fields are supposed to be negligible. This is also in accordance with the general believe that electric fields are not of importance in astrophysics, as mentioned in the Introduction.
In general, electrovacuum axisymmetric static spacetimes are considered  either with electric, magnetic or both fields
\cite{solutions}; no restriction needs to be imposed in this case. In the subclass of conformastatic fields, it is also possible to consider electric and magnetic fields simultaneously \citep{gutierrez2015variational}; nevertheless, the physical interpretation of the solutions in this case is more involved.
In reference \citep{gutierrez2015exact}, for instance,  this  kind  of  solutions has been  interpreted  in terms of magnetized halos around thin  disks.
As we will see below, the restriction to only the magnetic component of the four-potential allows us to describe the gravitational field of compact objects endowed with magnetic fields.

In fact, the assumption that only $A_{\varphi}$ is non-vanishing drastically reduces the complexity of the Einstein-Maxwell  equations (\ref{eq:E-Mequations}), which
 can be  equivalently written  as
\begin{eqnarray}
 & & \nabla^2 \phi     = \nabla \phi \cdot  \nabla \phi,\label{eq:EMexpli1}\\
 & & \phi_{,r}^{\,2}   = \frac{G }{c^4 r^2} e^{2\phi} A_{\varphi,z}^2,\\
  & &\phi_{,z}^{\,2}   = \frac{G }{c^4 r^2} e^{2\phi} A_{\varphi,r}^2,\\
  & &\phi_{,r}\phi_{,z}=-\frac{G }{c^4 r^2} e^{2\phi} A_{\varphi,r}A_{\varphi,z},\\
 & & \nabla \cdot (r^{-2} e^{2\phi} \nabla{A_{\varphi}})=0,
\end{eqnarray}
where $\nabla$  denotes  the usual gradient operator in  cylindrical  coordinates, and a comma indicates partial differentiation with respect to the corresponding variable. The symmetry properties of the above differential equations allows us to reduce them to a single equation. In fact,
by  supposing a solution in the  functional  form $\phi=\phi[U(r,z)]$,  where
$U(r,z)$ is  an arbitrary harmonic  function restricted  by  the  condition $U < 1$ for all $r$  and $z$ (\cite{PhysRevD.87.044010}, \cite{GRG2015ACG-P2015}), it is  not  difficult  to prove that
\begin{equation}
 \phi=-\ln{(1 - U)}, \quad
  A_{\varphi}(r,z) = \frac{c^2}{G^{1/2}}{\int_0^r{\tilde{r} U({\tilde{r},z}}}) d \tilde{r},
 \label{eq:solution}
\end{equation}
 represent a solution  of  the system (\ref{eq:EMexpli1}).

The electromagnetic field is pure magnetic. This  can be demonstrated by analyzing the electromagnetic invariant ${\cal F}=F^{\alpha\beta} F_{\alpha\beta} $, which in this case has the  form
\begin{equation}
{\cal F} \equiv F^{\alpha\beta} F_{\alpha\beta} = \frac{2 c^4 (U^2_{,r} + U^2_{,z})}{G (1 -U)^4 } \geq 0.
\label{eq:electromageticinvariant}
\end{equation}
In fact, the  nonzero  components  of  the  electromagnetic  field are
\begin{equation}
B_r= \frac{c^2}{G^{1/2}}rU_{,r} \quad \text {and} \quad
B_z= \frac{c^2}{G^{1/2}}rU_{,z} \ .
\label{eq:magneticfield}
\end{equation}
We conclude that any harmonic function $U(r,z)<1$ can be used to construct an exact conformastatic solution of the Einstein-Maxwell equations. This is an interesting result  which allows us to investigate the physical properties of  concrete  conformastatic spacetimes. Indeed, in the sections below we will investigate, as a particular example, one of the simplest harmonic solutions which turns out to describe the gravitational field of a punctual mass.

We now analyze the geodesic equations for a general case.
The  motion  of  a test  particle of mass $m$ and  charge $q$ moving  in a   conformastatic spacetime  given  by  the  line element (\ref{eq:metric})
is  described  by the Lagrangian
\begin{equation}
  {\cal L} =\frac{1}{2}m g_{\alpha\beta}  \dot{x}^{\alpha}\dot{x}^{\beta} + \frac{q}{c}A_{\alpha}\dot{x}^{\alpha},
  \label{eq:lagrangian}
\end{equation}
where  a  dot  represents differentiation with  respect to the  proper time.
The equations  of motion of the test particle  can  be  derived  from  the Lagrangian (\ref{eq:lagrangian}) by  using  the  Hamilton equations
\begin{eqnarray}
\dot{p}_{\alpha}=  - \frac{\partial \cal{H} }{\partial {x}^{\alpha}}, \quad
\dot{x}_{\alpha}=\frac{\partial \cal{H} }{\partial {p}^{\alpha}},   \nonumber \\
{\cal H} = \dot{x}^{\alpha} p_{\alpha} - {\cal L}\ , \quad
 p_{\alpha}=  \frac{\partial \cal{L} }{\partial \dot{x}^{\alpha}},
\label{eq: motionequations}
\end{eqnarray}
where $p_{\alpha}$ and $\cal H$ are the momentum and  Hamiltonian of  the  particle, respectively.
The coupled differential equations (\ref{eq: motionequations})  for  the  Lagrangian (\ref{eq:lagrangian}) are very
difficult  to  solve directly  by using an analytic approach. However, we can use the symmetry properties of the conformastatic field to find first integrals of the motion equations which reduce the number of independent equations. This is the approach we will use below.

Since  the  Lagrangian (\ref{eq:lagrangian}) does not  depend  explicitly  on the  variables $t$ and $\varphi$,  one can obtain the  following  two conserved  quantities
 \begin{equation}
   p_t= -mc e^{2\phi}\dot{t} \equiv -\frac{E}{c},
   \label{eq:energy}
\end{equation}
and
\begin{equation}
   p_{\varphi}= m r^2 e^{-2\phi}\dot{\varphi} + \frac{q}c{}A_{\varphi} \equiv L,
   \label{eq:angularmomentum}
\end{equation}
where $E$ and $L$ are, respectively,  the  energy and the  angular  momentum  of  the particle  as  measured by  an observer at rest at infinity.
Furthermore, the momentum $p_{\alpha}$ of  the  particle  can be  normalized  so that $ g_{\alpha\beta} \dot x^{\alpha} \dot x^{\beta} = - \Sigma$. Accordingly, for
the  metric (\ref{eq:metric}) we have
\begin{equation}
 -{c^2e^{2\phi} \dot{t}^2} + e^{-2\phi}(\dot{r}^2  + \dot{z}^2 + r^2 \dot{\varphi}^2  ) =- \Sigma,
 \label{eq:circuprecess}
\end{equation}
where $\Sigma = 0, c^2, -c^2$ for null,  time-like  and  space-like curves, respectively.

The  relations  (\ref{eq:energy}),  (\ref{eq:angularmomentum})  and (\ref{eq:circuprecess}) give  us  three  linear differential equations,
involving  the four  unknowns $\dot{x}^{\alpha}$. It is  possible to  study the motion of  test  particles  with only  these  relations, if we limit ourselves
to the  particular  case  of    equatorial trajectories, i. e.    $z = 0$. Indeed, since the gravitational configuration is symmetric with respect to the equatorial plane, a particle with initial state $z=0$ and $\dot z=0$ will remain confined to the equatorial plane which is, therefore, a geodesic plane.
Substituting the  conserved quantities (\ref{eq:energy}) and (\ref{eq:angularmomentum}) into Eq.(\ref{eq:circuprecess}), we  find
\begin{equation}
  \dot{r}^2 + \Phi = \frac{E^2}{m^2c^2}, \label{eq:motion}
\end{equation}
where
\begin{equation}
  \Phi(r) \equiv \frac{L^2}{m^2r^2}\left(1 - \frac{qA_{\varphi}}{Lc}  \right)^2 e^{4\phi} + \Sigma e^{2\phi}
  \label{eq:effectivepot}
\end{equation}
is an effective  potential.

\section{Circular  orbits}\label{sec:The  circular  orbits}
The    motion  of  charged test particles  is  governed  by  the  behavior of  the  effective potential (\ref{eq:effectivepot}). The  radius of  circular  orbits and  the corresponding values of  the
energy $E$ and  angular momentum  $L$ are  given by  the  extrema of the  function $\Phi$. Therefore, the  conditions  for  the  occurrence of  circular  orbits  are
\begin{equation}
  \frac{d\Phi}{dr} =0, \quad  \Phi= \frac{E^2}{m^2c^2}.
  \label{eq:circularconditios}
\end{equation}
We assume the convention that the  positive  value  of  the  energy  corresponds to  the positivity of the solution $E_{\pm} = \pm m c\Phi^{1/2}$.
Consequently, $E_{+} = - E_{-} =  m c\Phi^{1/2}$.

Calculating the condition (\ref{eq:circularconditios})  for  the  effective  potential   (\ref{eq:effectivepot}),  we  find the angular momentum  of the particle in   circular  motion
\begin{equation}
  L_{c \pm} =  \frac{q A_{\varphi}}{c}
                  +   \frac{q r A_{\varphi,r} e^{\phi} \pm \sqrt{ \left(q r A_{\varphi,r} e^{\phi}\right)^2
                  -   4 \Sigma c^2m^2r^3 \phi_{,r} \left(2 r \phi_{,r} - 1\right)} }
                     {2 c  e^{\phi} \left(2 r \phi_{,r} - 1\right)}.   \label{eq:circangularmomentum}
\end{equation}
Conventionally,  we  can associate  the plus and minus signs in the subscript of the notation $L_{c\pm}$ to  dextrorotation and  levorotation, respectively.

Furthermore, by  inserting the  value  of  the  angular momentum
(\ref{eq:circangularmomentum}) into the  second  equation of Eq.(\ref{eq:circularconditios}), we  obtain the  energy  $E_{\quad c  \pm}^{(\pm)}$ of  the particle  in a  circular  orbit as
\begin{equation}
   E_{\quad c  \pm}^{(\pm)} = \pm\, m c e^{\phi} \left(\Sigma + \xi^{(\pm)}_{\quad c}\right)^{1/2}
   \label{eq:energycir},
\end{equation}
where
\begin{equation}
    \xi^{(\pm)}_{\quad c}= \frac{\left[ q r A_{\varphi,r} e^{\phi} \pm \sqrt{ \left(q r A_{\varphi,r}
    e^{\phi}\right)^2 - 4 \Sigma c^2 m^2 r^3   \phi_{,r} \left(2 r \phi_{,r} - 1\right)} \right]^2 }
                            {4 m^2 c^2 r^2   \left(2 r \phi_{,r} - 1\right)^2}.
                            \label{eq:xi-general}
\end{equation}
Therefore,  each  sign of the value of  the  energy corresponds to two kinds of  motion (dextrorotation and  levorotation) indicated  in  (\ref{eq:energycir})  and (\ref{eq:xi-general})
by  the  superscripts $(\pm)$.

An  interesting particular  orbit  is  that  one  in which  the  particle is  located  at  rest   $(r_r)$ as  seen  by an observer at  infinity,
i.e. $L=0$. These orbits are  therefore characterized  by  the  conditions
\begin{equation}
 L=0, \quad \frac{d \Phi}{dr} =0.
 \end{equation}
For  the metric (\ref{eq:metric})  these  conditions give  us  the  following  equation
\begin{equation}
\frac{2 e^{2 \phi}}{ m^2 c^2 r^3 } \left[  q^2 A_{\varphi}  \left(2 r  A_{\varphi} \phi_{,r}
            + r  A_{\varphi, r} -  A_{\varphi} \right) e^{2 \phi}
            + \Sigma  m^2c^2  r^3 \phi_{,r}\right] =0\ .
						\label{eq:rrest}
\end{equation}
To find the value of the rest radius $r_r$, we  must   solve  Eq.(\ref{eq:rrest}).

Notice  that from Eqs. (\ref{eq:rrest}) and (\ref{eq:energycir}) it follows that   if  $ e^{2 \phi}=0$ for  an  orbit
with a  rest radius $r=r_r$, the  energy  of  the  particle  is  $E_r= 0$.
In the  case $ e^{2 \phi}
\neq 0$,  we  have
\begin{equation}
 E_{\quad r  \pm}^{(\pm)} = \pm m c e^{\phi} \left(\Sigma + \xi^{(\pm)}_r \right)^{1/2}
 \label{eq:energyrest},
\end{equation}
where
\begin{equation}
\xi^{(\pm)}_{\quad r} =\frac{q^2 e^{2\phi}   \left[  r A_{\varphi, r} \pm \left(  r A_{\varphi, r}
+ 2 A_{\varphi}  \left(2 r \phi_{,r} - 1\right)   \right)  \right]^2}
{4 m^2 c^2 r^2   \left(2 r \phi_{,r} - 1\right)^2}.
\end{equation}
This analysis indicates that it is possible to have a test particle at rest with zero angular momentum ($L=0)$ and non-zero energy $(E_r\neq 0)$.  This is a non-trivial effect that in the case of vanishing magnetic field has been associated with the existence of repulsive gravitational effects in Einstein gravity
\citep{PhysRevD.83.104052}.

The  minimum radius  for  a  stable circular  orbit  corresponds to an inflection  point of  the  effective potential function. Thus, we  must  solve  the  equation
\begin{equation}
  \frac{d^2\Phi}{dr^2} =0,
  \label{eq:stableconditions}
\end{equation}
under the condition that the angular momentum is given by Eq.(\ref{eq:circangularmomentum}).  From  Eqs.(\ref{eq:circularconditios}) and (\ref{eq:stableconditions}), we  find  that  the  radius and   angular momentum of the last stable circular orbit are  related by  the  following
equations
\begin{eqnarray}
\frac{2 e^{4 \phi}}{m^2 c^2 r^4} \Big[
                                 (Lc - q A_{\varphi})^2 \left(  8 r^2 \phi_{,r}^2
                                 +  2r^2 \phi_{,rr}  - 8 r \phi_{,r }  + 3\right)
                              \nonumber \\
                              + (Lc - q A_{\varphi})  \left( -8 q r^2  A_{\varphi, r} \phi_{,r}
                              - qr^2 A_{\varphi, r r }  + 4q r A_{\varphi, r} \right)
                              \nonumber \\
                             + q^2 r^2 A_{\varphi, r} ^2 e^{2\phi} + 2c^2\Sigma m^2 r^4 \phi_{,r}^2
                             + c^2\Sigma m^2r^4 \phi_{,rr} \Big]     = 0,  \label{eq:secondderstable}
\end{eqnarray}
and
\begin{eqnarray}
\frac{2 e^{2 \phi}}{m^2 c^2 r^3}
            \Big[     e^{2\phi}(Lc - q A_{\varphi}) \Big( (Lc - q A_{\varphi})(2r\phi_{,r}  - 1)
                       - qr A_{\varphi, r} \Big) \nonumber \\
                        +\;  \Sigma m^2 c^2 r^3 \phi_{,r}   \Big]   = 0.
                         \label{eq:firstderstable}
  \end{eqnarray}
It is  possible  to  solve Eq.(\ref{eq:firstderstable})  with respect to the  stable circular
orbit  radius  which then becomes a  function of  the  free  parameter $L$.  Alternatively, from Eq. (\ref{eq:secondderstable} )
we find the  expression
\begin{eqnarray}
L^{\pm}_{lsco}&=&\frac{q A_{\varphi}}{c} +
                         \Bigg\{   q e^{\phi} \left(  8r^2 A_{\varphi,r} \phi_{,r} ^2
                         + r^2  A_{\varphi ,rr}    - 4 r A_{\varphi ,r} \right)
                         \nonumber\\
                        &&  \pm  \Bigg[ q^2 e^{2\phi} \left(  8r^2 A_{\varphi,r} \phi_{,r} ^2
                        + r^2  A_{\varphi ,rr}    - 4 r A_{\varphi ,r} \right)^2
                         \nonumber\\
                         &&- 4 ( q^2 r^2 A_{\varphi,r} ^2 e^{2\phi} + 2c^2\Sigma m^2r^4\phi_{,r}^2
                         + c^2\Sigma m^2 r^4 \phi_{,rr})
                         \nonumber\\
                         &&\times (8 r^2 \phi_{,r}^2   + 2r^2 \phi_{,rr}  - 8r\phi_{,r} + 3)
                         \Bigg]^{1/2}  \Bigg\}
                          \nonumber\\
                          && \times \left[ 2 c  e^{\phi} (8 r^2 \phi_{,r}^2   + 2r^2 \phi_{,rr}
                          - 8r\phi_{,r} + 3)    \right]^{-1}
                         \label{eq:angulmomentstableosrbit}
\end{eqnarray}
for  the  angular    momentum  of  the last stable circular  orbit. Equation (\ref{eq:angulmomentstableosrbit}) can then be  substituted  in Eq. (\ref{eq:firstderstable}) to find  the  radius  of  the last  stable circular   orbit.

In this section, we found the expressions for the physical quantities which characterize the behavior of a charged test particle, moving along a circular trajectory in the gravitational field of a conformastatic mass distribution endowed with a magnetic field. These results are completely general, and can be applied to any solution of the corresponding Einstein-Maxwell equations.

\section{The  field  of  a  punctual mass in Einstein-Maxwell  gravity}
        \label{The  circular motion   in the  field  of  a  punctual mass in Einstein-Maxwell  gravity}

We  now   illustrate  the  results  obtained in the  precedent  section, focusing on the main physical properties of test particles moving along circular orbits.
As shown before, the class of harmonic conformastatic solutions is of particular interest, because all the metric components and the magnetic field are defined in terms of a single harmonic function. Let us consider one of the simplest harmonic functions which in Newtonian gravity would describe the gravitational field of a punctual mass, namely,
\begin{equation}
    U(r,z)=-\frac{G M }{c^2 R},  \quad R^2= r^2 + z^2\ ,
    \label{eq:Upotential}
\end{equation}
where $M$ is a real constant.  According to Eq.(\ref{eq:solution}), for  the  metric  and  electromagnetic  potentials  we have
\begin{equation}
\phi(r, z)= - \ln{  \left(1 + \frac{G M}{c^2 R}  \right) } \ ,
\label{eq:explicsolgr}
\end{equation}
and
\begin{equation}
A_{\varphi}(r, z)= \sqrt{G} M \left( 1 - \frac{z}{R} \right) \ ,
\label{eq:explicsolmag}
\end{equation}
respectively.
At spatial infinity, the magnetic potential is non-zero and constant, except at the symmetry axis where it vanishes. Notice also that on the equatorial plane the magnetic potential is constant everywhere. As for the metric potential, its physical significance can be investigated by considering the asymptotic  behavior  of  the
metric  component $g_{tt}$ for which we obtain
\begin{equation}
 \lim_{R  \to\infty} g_{tt}(r,z) \approx -1 + \frac{2 G M}{c^2 R}  -  \frac{3 G^2 M^2}{c^4R^2} +
                {\cal O}\left(\frac{1}{R^3}\right)\ .
\label{eq:asymptoticbehaviour}
\end{equation}
Accordingly, this particular solution can be interpreted as describing the gravitational field of a punctual mass on the background of a magnetic field. In the limiting case $M\rightarrow 0$, we obtain the Minkowski spacetime, indicating that $M$ is the source of the gravitational and the magnetic field as well. This result can be corroborated by analyzing the Kretschmann scalar   ${\cal K}= R_{\alpha\beta\gamma\delta} R^{\alpha\beta\gamma\delta}$ and  the electromagnetic invariant $\cal F$
(see Eq.(\ref{eq:electromageticinvariant})) which in this case have the  following  expressions
\begin{equation}
{\cal K}= \frac{8 M^2 c^8 G^2  (G^2M^2 + 6c^4R^2)}{(c^2R + GM)^8},
\label{eq:kretschmann}
\end{equation}
and
\begin{equation}
{\cal F} = \frac{2 G M^2 c^8}{(c^2 R + G M)^4},
\label{eq:electromageticinvariant-part}
\end{equation}
respectively.  In  addition, from the expressions (\ref{eq:asymptoticbehaviour}), ( \ref{eq:kretschmann})  and
(\ref{eq:electromageticinvariant-part}), we  conclude  that the  gravitational  field  is asymptotically Schwarzschild-like and
singularity-free.

\begin{figure}
	\includegraphics[width=0.4\columnwidth]{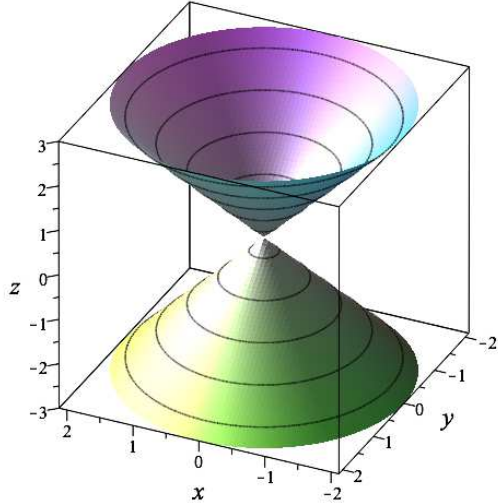}
        \caption{Illustration of the spatial distribution of the magnetic lines of force of a punctual conformastatic source located at the origin of coordinates.}
    \label{fig0}
\end{figure}

The field lines of the magnetic field are given by the ordinary differential equation
$
{dr}{B_z}= {dz}{B_r}.
$
Moreover, the  nonzero  components  of  the  magnetic  field  are $B_r =  G^{1/2} M r^2 R^{-3}$ and $B_z = G^{1/2} M r z  R^{-3}$ . Thus,  we  see that
the equation
\begin{equation}
z^2 = \frac{\left(1  - \gamma \right)^2}{\gamma (2 - \gamma)}r^2\ ,
\label{eq:linesofforce}
\end{equation}
where $ 0 < \gamma < 2$,  represents the lines of force of the magnetic field. The components of the magnetic field vanish at spatial infinity, and diverge at the origin $R=0$. This, however, is not a true singularity as can be seen from the expression for the electromagnetic invariant (\ref{eq:electromageticinvariant-part}).
In Fig.\ref{fig0}, we illustrate the spatial behavior of the lines of force of this magnetic configuration.
It shows that the source of the magnetic field coincides with the punctual mass, in accordance with the analytic expressions for the gravitational and magnetic potentials.

\subsection{Circular  motion  of  a    charged  test particle}
               \label{sec:The  circular  motion  of  a  test  charged  particle in  a  particular  effective potential}
Consider  the case  of  a  charged  particle moving  in the  conformastatic field of a punctual mass  given   by Eqs. (\ref{eq:explicsolgr}) and  (\ref{eq:explicsolmag}).
This  means  that  we  are considering  the  motion  described  by  the  following  effective  potential
\begin{equation}
 \Phi(r)= \frac{c^6 r^2 (Lc - q\sqrt{G}M)^2}{m^2 (c^2r + G M)^4}
           + \frac{\Sigma c^4 r^2}{(c^2r + G M)^2}.
           \label{eq:effectivepotpart}
\end{equation}
We  note  that
\begin{equation}
e^{\phi(r)}=\frac{c^2 r}{c^2 r + G M}
\end{equation}
on the equatorial plane.

\begin{figure}
	\includegraphics[width=0.4\columnwidth]{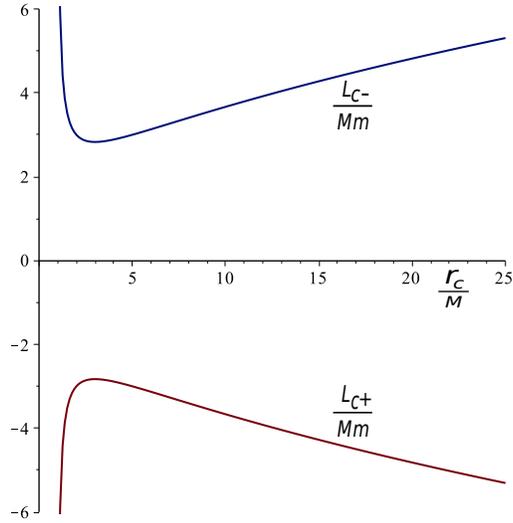}
        \caption{Angular momentum of a neutral test particle in terms of the radius orbit $r_c/M$.}
    \label{fig1}
\end{figure}

According to the general results of the previous section, the  angular momentum and  the  energy for  a  circular  orbit with radius $r_c$ are  given  by
\begin{equation}
 L_{c\;\pm} = \frac{q \sqrt{G}M}{c} \mp  \frac{(c^2 r_c + GM)m}{c^2}
                     \sqrt{ \frac{\Sigma G M}{c^2 r_c - GM}}
                       \label{eq:Lco_p}
\end{equation}
and
\begin{equation}
     E_{c\;\pm} =\pm \frac{ m c^4  }
                         { \left(c^2  r_c + GM \right)} \sqrt{\frac{\Sigma r_c^{3}}{{c^2} r_c - {G M}} }
                         \label{eq:Eco_p}
\end{equation}
respectively. From  Eqs.(\ref{eq:Lco_p}) and (\ref{eq:Eco_p}), we  conclude   that  in  order  to  have   a  time-like  circular  orbit the  charged particle must  be placed at  a  radius $r_c > GM/c^2$.  In Fig.\ref{fig1}, we illustrate the behavior of the angular momentum for the particular case of a neutral $(q=0)$ particle.
We see that $L_{c+}$ ($L_{c-}$) is always negative (positive) for all allowed values  $ r_c> GM/c^2$, and diverges in the limiting case  $ r_c=GM/c^2$. Since the charge $q$  enters the angular momentum (\ref{eq:Lco_p}) as an additive constant, it does not affect the essential behavior of $L_{c\pm}$, but it only moves the curve along the vertical axis.  However, the value of the effective charge $q/m$  can always be chosen in such a way that either $L_{c+}$ or $L_{c-}$ become zero at a particular radius $r_r$. For instance, for $L_{c+}$ to become zero, the charge $q$ must be positive and greater than a certain value.
This is the first indication that a circular orbit with zero angular momentum occurs as the result of the electromagnetic interaction between the particle electric charge and the magnetic field of the punctual source.

\begin{figure}
	\includegraphics[width=0.4\columnwidth]{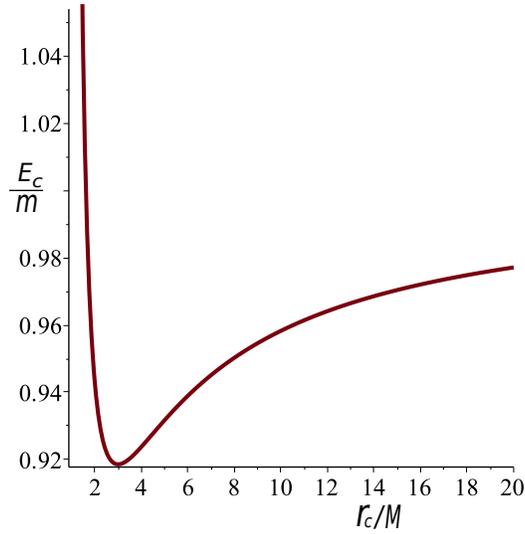}
        \caption{Energy of a charged  test particle in terms of the radius orbit $r_c/M$.}
    \label{fig2}
\end{figure}

As for the energy of circular orbits, we see from Eq.(\ref{eq:Eco_p}) that it does not depend explicitly on the value of the charge,
but only on the radial distance from the central punctual mass. This, however, does not mean that the energy does not depend on the charge at all. Indeed,
from Eq.(\ref{eq:Lco_p}) we see that, for a given angular momentum, the charge $q$ influences the value of the circular orbit radius which, in turn, enters the expression for the energy.
The behavior of the energy in terms of the radial distance is depicted in Fig.\ref{fig2}.
As the radius approaches the limiting value $r_c/M= G/c^2 = 0.774 \times 10^{-28}$ cm\;g$^{-1}$, the energy diverges indicating that a test particle cannot
be situated on the minimum radius.
The energy has an extremal located at the radius $r_c/M = 3G/c^2 = 2.227 \times 10^{-28}$ cm\;g$^{-1}$. We will see below that it corresponds to a particular
orbit at which the particle stays at rest.
At spatial infinity, we see that $E_{c+} \rightarrow mc^2$ which corresponds to  the rest energy of the particle outside the influence of the magnetic and gravitational fields.

\begin{figure}
	\includegraphics[width=0.4\columnwidth]{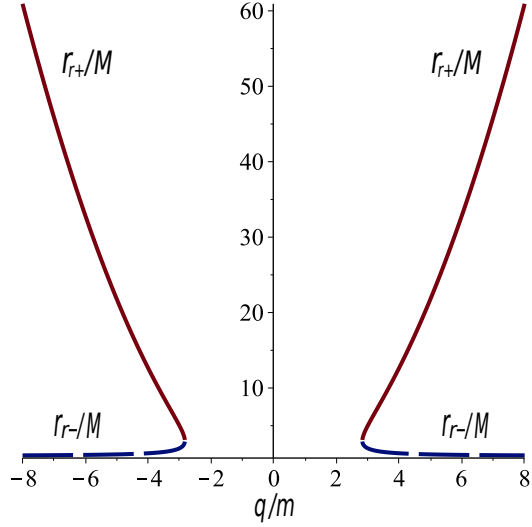}
        \caption{Radii  of  the  time-like  orbits characterized  by  the  conditions
        $L=0$ and $d\Phi/dr=0$ (see Eq.\ref{eq:rrestp}). In this graphic  the  radii
        $r_{r +}/M$ (solid  curve) and   $r_{r -}/M$ (dashed curve)    are plotted  as functions
        of  $q/m$.}
    \label{fig:EffectPot}
\end{figure}

Let us now  consider the conditions under which the particle can remain at rest $(L=0)$ with respect to an observer at infinity. From Eq.(\ref{eq:rrest})  for  $q\neq 0$  we  find  the  trivial   rest radius $r_r=0$ for which the  energy of the  particle is $E_r=0$. In addition, the non-trivial solution is given by the rest  radii
\begin{equation}
 \frac{r_{r \; \pm}}{M}= \frac{ c^2 q^2  - 2 \Sigma G m^2 \pm  \sqrt{ c^2q^2\left(c^2q^2
                                  - 8 \Sigma G m^2\right)} }{2 \Sigma m^2 c^2}.
 \label{eq:rrestp}
\end{equation}
The behavior of these radii is depicted in Fig. \ref{fig:EffectPot}. We can see that these solutions are physically realizable in the sense that the radii are always positive for all values of $q/m$ that satisfy the condition $q^2 \geq 8 m^2 G$.
Indeed, the existence of a  rest radius is   restricted  by the  discriminant $q^2 - 8 m^2 G$ in Eq.(\ref{eq:rrestp}).
For time-like test particles with  $q^2/m^2> 8G$, there exist an inner radius $r_{r-}$ and an outer radius
$r_{r+}$  at which the particle can remain at rest.  For the limiting value $q^2/m^2=8G$, the two radii coincide with $r_{r+}=r_{r-}=3GM/c^2$.
 Instead, if  $q^2/m^2<8G$, no radius exists at which the particle could stay at rest. Clearly, the existence of a rest radius is determined by the value of the test particle effective charge, indicating that a zero angular momentum orbit is the consequence of an electromagnetic effect due to the interaction of the test charge and the magnetic background.

\begin{figure}
	\includegraphics[width=0.4\columnwidth]{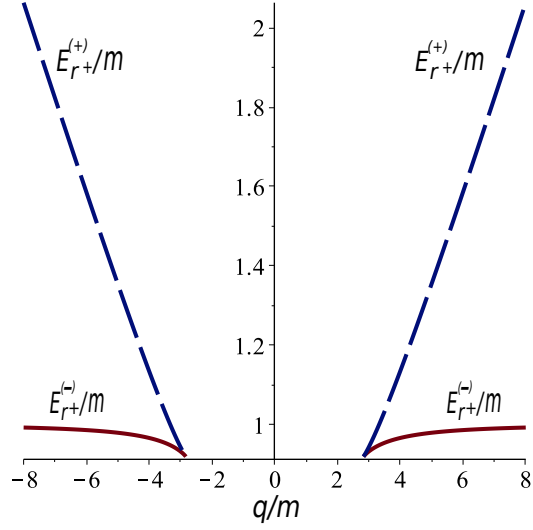}
        \caption{Energy of  the  charged  particles for  the  time-like  orbits characterized  by  the  conditions
        $L=0$ and $d\Phi/dr=0$ (see Eqs.(\ref{eq:restEnergya}) and (\ref{eq:restEnergyb})). In this graphic  the  energy
        $E_{r +}/m$ for  the  radius  $r_{r+}$(solid  curve) and   $E_{r +}/m$ for  the  radius $r_{r-}$(dashed curve)     are plotted as functions
        of  $q/m$.}
    \label{fig:restEnergyp}
\end{figure}

For  the  outer radius  $r_{r\,+}$  we  have two possible positive values of the corresponding energy, namely,
\begin{equation}
 E_{r +}^{(\pm)} = +  \frac{m c^2 \sqrt{2 q^2 \left[q^2  -  2 G m^2  \pm \sqrt{ q^2 (q^2 - 8 G m^2 )^2}  \right]^3}}{\left[q^2   \pm \sqrt{q^2 (q^2  - 8 Gm^2)^2}  \right]^2} ,
  \label{eq:restEnergya}
 \end{equation}
 whereas   for  the  inner radius $r_{r\,-}$,  the  two possible energies are always  negative
\begin{equation}
 E_{r -} ^{(\pm)}= - E_{r +}^{(\pm)} \ .
\label{eq:restEnergyb}
\end{equation}
In Fig. \ref{fig:restEnergyp}, we show the behavior of the energy at the outer rest radius. The minimum value of $\pm (3\sqrt{6}/8) m c^2$ is reached for $q^2 = 8 G m^2$, i.e., when the inner and outer radii coincide.

\begin{figure}
	\includegraphics[width=0.4\columnwidth]{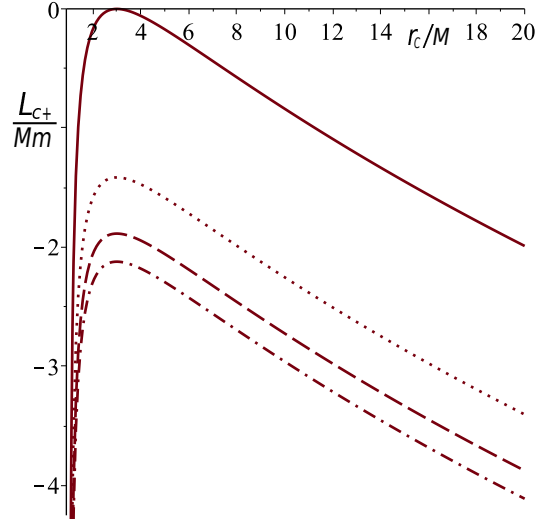}
        \caption{Angular  momentum  of  charged  particles in  a  time-like circular  orbit (see Eq. (\ref{eq:Lco_p})). In this graphic
         the angular  momentum $L_{c }^{+}/Mm$ is plotted as a function of the  radius   $r_{c}/M$ for  some  values  of $q/m$. The  continuous  curve
         corresponds  to  the  value $q/m=2\sqrt{2}$. 				
				}
    \label{fig:AngularMomentumCircularp}
\end{figure}

\begin{figure}
	\includegraphics[width=0.4\columnwidth]{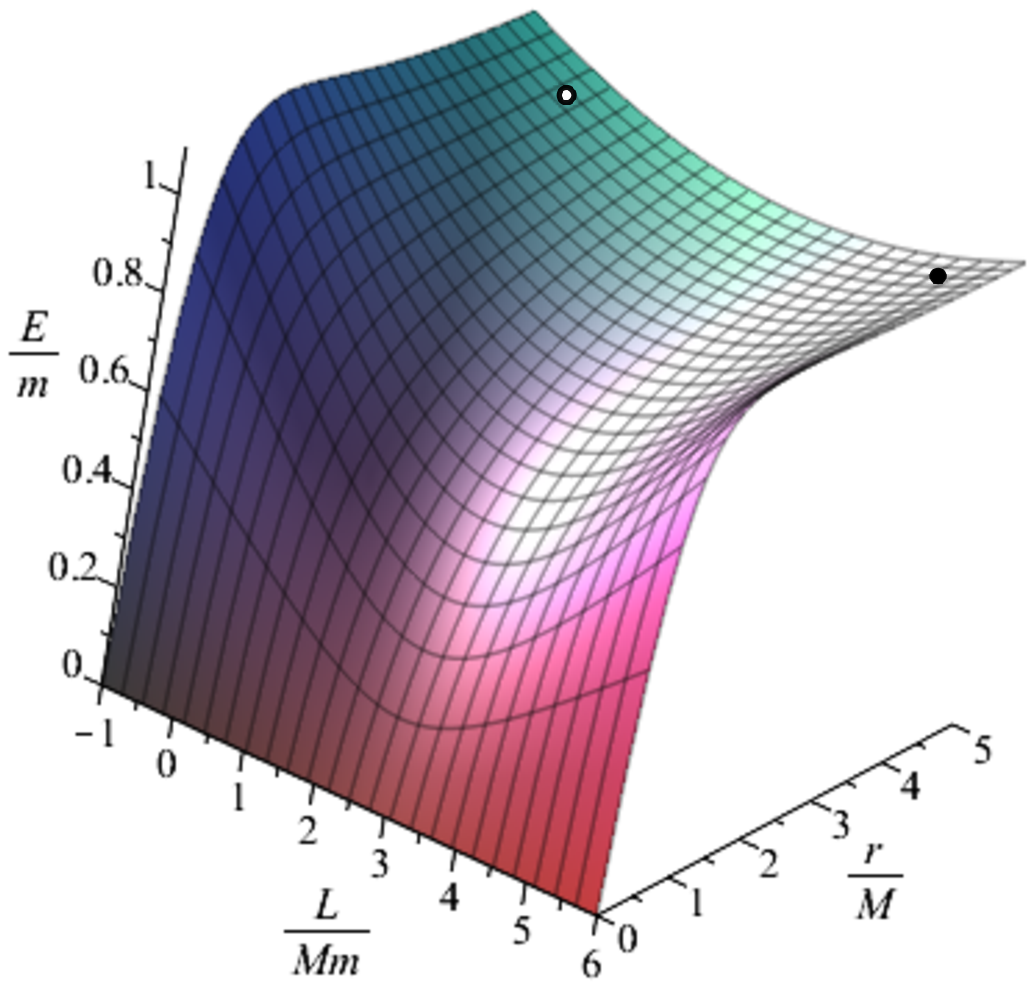}
        \caption{The energy   of  a  particle  with  charge-to-mass-ratio $q/m=2\sqrt{2}$ is  plotted as  a  function  of $r/M$  and  the  angular momentum $L/Mm$.
        Time-like circular orbits  exist for  $r/M > 1$. The  last stable circular orbit is  represented  by  a  black  point ($r/M = 3,
				L_{lsco+}=4\sqrt{2}Mm $).
        The  orbit  with the same radius ($r/M=3$), but zero angular momentum ($L_{lsco-}=0$)  is represented  by a white point.}
    \label{fig:Energyplot}
\end{figure}

We now investigate the properties of the last stable circular orbit. According to Eq.(\ref{eq:angulmomentstableosrbit}), the angular momentum for this particular orbit must satisfy the relationship
\begin{equation}
 L_{lsco\;\pm} = \frac{q \sqrt{G}M}{c} \pm  \frac{(c^2 r + GM)m}{c^2}
                     \sqrt{ \frac{\Sigma G M ( 2c^2 r - GM)}{G^2 M^2  - 6 c^2  GM r + 3c^4r^2}}.
                       \label{eq:Llsco_p}
\end{equation}
On the other hand, the angular momentum for any circular orbit is given by Eq.(\ref{eq:Lco_p}). Then, the comparison of  Eqs.(\ref{eq:Llsco_p}) and (\ref{eq:Lco_p}) yields the condition $r (c^2 r - 3 GM) =0$. Therefore, the  radius of the last  stable circular  orbit  is given by
\begin{equation}
r_{lsco}=\frac{3GM}{c^2} \ ,
\label{eq:stableratio}
\end{equation}
which, remarkably, does not depend on the value of the charge $q$.
 The corresponding angular momentum can be expressed as
\begin{equation}
 L_{lsco\;\pm} = \frac{q \sqrt{G}M}{c} \pm
                          \frac{  2\sqrt{2 \Sigma } G M m} {c^2}
                       \label{eq:Llsco_p3GM},
\end{equation}
and the  energy reduces to
\begin{equation}
 E_{lsco\;\pm} = \pm \frac{3}{4} \sqrt{\frac{3\Sigma}{2}} m c \ .
                      \label{eq:Elsco_p3GM}
\end{equation}
As  we  can see, the  angular momentum depends explicitly on the value of the  mass  $m$ and  charge $q$ of  the test  particle.
Additionally, for space-like  curves   the  angular momentum of
the last  stable circular  orbit is  not  defined,  whereas  for  a  null  curve it is  $L_{lsco\;\pm} = {q \sqrt{G}M}/{c}$, and
for a time-like particle it is $ L_{lsco\;\pm} = (q \sqrt{G}  \pm   2\sqrt{2  } G m)M /c$. Accordingly,   if  the  charge
of  the  particle is $q=2\sqrt{2G}m$,  then $L_{lsco-}=L_{c+}=L_{r}=0$. Analogously, if $q=-2\sqrt{2G}m$,  then $L_{lsco+}=L_{c-}=L_{r}=0$.

Thus, we  conclude  that the  last  stable  circular  orbit
occurs  at the radius   $r = 3 GM/c^2$, independently of the value  of  the charge.
Moreover,  on the  last  stable  orbit   the  particle  is at  rest,    if  the value of  the  charge is
$q=\pm2\sqrt{2G}m$ (see Fig. \ref{fig:AngularMomentumCircularp}).

In Fig. \ref{fig:Energyplot}, we present the general behavior of the energy function $E=\pm m c \sqrt{\Phi}$ in terms of the effective potential  $\Phi$ given  by
Eq.(\ref{eq:effectivepotpart}). The branch corresponding  to the positive energy   of  a  particle  with  charge-to-mass-ratio $q/m=2\sqrt{2G}$ is  plotted as  a  function  of $r/M$  and  the angular
momentum $L/Mm$.
We see that the effective potential (energy)  tends to a  constant at  infinity.  Since  the  radius  for  the  last stable  circular  orbit  is $r_{lsco}= 3GM/c^2$,
for a particle with charge $q/m=2\sqrt{2G}$ it is possible to stay at rest with $L_{lsco-}=0$ (white point in  Fig.\ref{fig:Energyplot}) or with angular momentum
$L_{lsco+} = 4\sqrt{2}GMm/c$ (black point in  Fig.\ref{fig:Energyplot}). A  similar  result is obtained  if the  charge  is  negative, corresponding to the angular  momentum  $-4\sqrt{2}G/c$, which indicates  rotation in the counterclockwise direction.

It  is  worth  noticing  that a  neutral  particle cannot stay at  rest with  zero angular momentum. This  can be  deduced  by replacing  $q=0$ in Eqs.(\ref{eq:rrestp}),
(\ref{eq:restEnergya}) and (\ref{eq:restEnergyb}). In fact, with a zero value of $q$ we  obtain a negative rest radius.
Finally, from Eqs.(\ref{eq:stableratio})  and (\ref{eq:Llsco_p3GM}) we  see  that  the  last  time-like stable  circular  orbit  for    neutral test   particles
can be placed at  $r = 3 GM/c^2$  with  angular  momentum   $L_{lsco\;\pm} =  {  \pm 2\sqrt{2 } G M m} /{c}$. Moreover, a  neutral massless test particle  can
get $L=0$ only at $r=0$, as  expected.

We conclude that from an observational point of view the most important characteristic of a conformastatic punctual mass is the radius of the last stable circular orbit $r_{lsco}=3GM/c^2$, which does
not depend on the value of the test charge $q$. For simplicity, consider the case $q=0$.  We should compare this value  of $r_{lsco}$ with the corresponding one in the case of axisymmetric fields. The
Reissner-Nordstr\"om (RN) metric with charge $Q$ would be the obvious candidate to perform the comparison. It is well-known that the magnetic RN metric can easily be obtained by changing 
$Q^2\rightarrow P^2$, where $P$ is the magnetic charge,  in all the components of the metric. If we then suppose that $P$ is proportional to $M$, we would obtain the corresponding metric for an axisymmetric punctual mass with a
magnetic field. In fact, this metric becomes spherically symmetric as a result of the symmetry properties of the field equations. The motion of a neutral test particle in the RN field was investigated
in detail in \citep{PhysRevD.83.104052}, where it was shown that $r_{lsco}^{_{RN}} \in [3/2,6) GM/c^2$ for all the relevant values of $Q/M$. This interval includes the value $3GM/c^2$ that corresponds
to the conformastatic case. This implies that by only measuring the radius of the last stable circular orbit, it is not possible to differentiate between axial and conformal symmetry.  We will
therefore consider a different test particle motion in the next section.

\section{Perihelion advance in a  conformastatic magnetized  spacetime}
    \label{sec:The perihelion advance in a  conformastatic magnetized  spacetime}
One of the most important tests of general relativity and modified theories of gravitation in astrophysical scale is the perihelion advance of celestial objects. In this section, we present
 the analytic expressions which determine the perihelion advance of charged test particles, moving in a conformastatic spacetime under the presence of a magnetic field.
Starting from the first integral (\ref{eq:circuprecess}),   we
restrict the analysis to the  motion of a particle on the
plane with $z=0$. Then,  we  have \begin{equation}
   \Bigg(\frac{dr}{d\varphi}\Bigg)^2 = - r^2 \Bigg[ 1  +  \frac{m^2r^2}{\big(L - \frac{q}{c}A_{\varphi}\big)^2}
                                   \Bigg(\Sigma (1- U)^2  - \frac{E^2 }{m^2 c^2} (1-U)^4 \Bigg)
    \Bigg],
   \label{eq:orbit}
\end{equation}
where all the  quantities are evaluated at $z=0$ and we  have  used  the  expressions  for the
energy  and  angular momentum of  the  particle  given  by Eqs.(\ref{eq:energy}) and
(\ref{eq:angularmomentum}), respectively.   This expression for the perihelion advance is valid for any
harmonic function $U=U(r,z=0)$.

To evaluate the perihelion advance in a concrete example, we consider the punctual solution with $U=\frac{GM}{c^2r}$
for a neutral test particle ($q=0)$. Then, introducing the auxiliary function $v= 1 + \frac{GM}{c^2r}$, Eq.(\ref{eq:orbit})
reduces to
\begin{equation}
\left(\frac{d v}{d\varphi}\right)^2 + (v-1)^2 = - \frac{G^2 m^2 M^2}{c^4 L^2} \left(c^2 v^2 -\frac{E^2}{m^2 c^2} v^4 \right)
\ .
\label{padv}
\end{equation}
Calculating the derivative of the above equation, we obtain
\begin{equation}
\frac{d^2 v}{d\varphi^2} + (1+\alpha^2) v = 1+ 2\beta^2 v^3\ ,\quad
\alpha =\frac{GmM}{cL}\ , \quad \beta= \frac{GME}{c^3 L} \ .
\label{padv1}
\end{equation}
The analytic solution to this equation can be found by using standard methods of ordinary differential equations. However, it is
enough to consider an approximate solution of the form
\begin{equation}
v = v_c(1+\omega) \ ,
\end{equation}
where $v_c= 1 + \frac{GM}{c^2r_c}$ is the solution for a circular orbit with radius $r_c$. Then, from Eq.(\ref{padv1}) we obtain
\begin{equation}
v_c \frac{d^2 \omega}{d\varphi^2} + (1+\alpha^2 - 6\beta^2)v_c \omega + (1+\alpha^2-2\beta^2)v_c  = 1\ ,
\label{padv2}
\end{equation}
where we have considered only linear terms in $\omega$. In the limiting case of a circular orbit ($\omega=0)$, the condition
$1+\alpha^2-2\beta^2= \frac{1}{v_c} \sim 1 - \frac{GM}{c^2r_c}$ must hold and, consequently, Eq(\ref{padv2}) reduces to
\begin{equation}
\frac{d^2 \omega}{d\varphi^2} + \left(1 - \frac{GM}{c^2 r_c}  - 4 \beta^2\right) \omega = 0 \ ,
\end{equation}
whose solution can be written as
\begin{equation}
\omega = \omega_0 \cos \left[\left(1 - \frac{GM}{c^2 r_c}  - 4 \beta^2\right)  \varphi \right]\ .
\end{equation}
The perihelion shift can be calculated at the maximum value of the solution, i.e., when $ \left(1 - \frac{GM}{c^2 r_c}  - 4 \beta^2\right)  \varphi = 2\pi$, which
can be written as $\phi =  2\pi + \Delta \varphi$, with $\Delta\varphi = \pi\left(\frac{GM}{c^2 r_c}  + 4 \beta^2\right)$. A straightforward computation by using
Eqs.(\ref{padv1}), (\ref{eq:Lco_p}) and (\ref{eq:Eco_p}), to the first order in $1/r_c$, leads to
\begin{equation}
\Delta \varphi = \frac{5\pi G M}{c^2 r_c} \ .
\end{equation}
This is the final value of the perihelion advance for a neutral test mass in the gravitational field of a punctual mass, endowed with a magnetic field. This compares with the value obtained for the
Schwarzschild spacetime $\Delta \varphi = \frac{6\pi G M}{c^2 r_c}$. The perihelion advance around a punctual magnetic mass is therefore always smaller than the value obtained in Einstein gravity
alone. We conclude that the perihelion advance permits us to differentiate between a spherically symmetric mass and a conformally symetric punctual magnetic mass.

\section{Conclusions}
\label{sec:conclusions}

In  this  work, we  have  shortly shown   the  characteristics of  the  motion of  a  charged particle  along  circular orbits in a  spacetime described  by   a  conformastatic solution  of the
Einstein-Maxwell  equations, which is also a solution of the general axisymmetric static electrovacuum Weyl class.

  As  a  particular  example  we  have  considered the  case  of  a  charged particle moving  in the gravitational  field   of  a  punctual  source placed at the origin of
coordinates. Our  analysis  is  based  on the  study of the behavior of an  effective  potential that  determines  the  position and  stability  properties  of  circular  orbits. We have found that a
classical  radius $r=3GM/c^2$ of circular  orbits  exists with  zero  angular  momentum.  This phenomenon is  interpreted as  a  consequence of  the repulsive electric  force that exists between the charge distribution and the charged test particle.
Interestingly, we have found this effect in a singularity-free spacetime, implying that it is not  exclusive to the  case  of  naked  singularities. Indeed, other configurations show a similar behavior even in the case of non-test particles;
for  instance, the the Majumdar-Papapetrou system \citep{maj47,pap47}, which is the subset of the Einstein-Maxwell-charged dust matter theory with the peculiar characteristic that the charge of each particle is equal to its mass.

Moreover, we  have  obtained  a region of stability
determined  by the angular momentum $L_{lsco}^{\pm}/Mm=q\sqrt{G}/(mc) + 2\sqrt{2}G/c$  and  the  radius $r=3GM/c^2$.
It  is  worth  noticing  that a  neutral  particle can not  be  located at  rest with  angular momentum  zero.  We also notice  that  the  last  time-like stable  circular  orbit  for    neutral test   particles  with  $m\neq  0$ can be placed  at  $r = 3 GM/c^2$  with  angular
momentum   $L_{lsco\;\pm} =  {  \pm 2\sqrt{2 } G M m} /{c}$,  and  that neutral massless particles can  get $L=0$  only when they   are  placed  at $r=0$, as
expected.

In addition,  we  have  also  calculated  an  expression  for    the  perihelion advance  of a test  particle in a general   magnetized conformastatic  spacetime.
In the particular case of a punctual magnetic mass, we found the explicit value of the perihelion advance, which turns out to be $5/6$ of the Schwarzschild value.

Our results can be used to test the possibility that conformally symetric mass distributions exist in Nature. Suppose that some observations show that the radius of
the last stable circular orbit around a static compact object of mass $M$ is $r_{lsco}= {3GM/c^2}$. Since this value is predicted for the punctual magnetic mass
as well as for a  Reissner-Nordstr\"om source, it is not possible to determine the symmetry of the source. Nevertheless, if observations show that the perihelion
shift around a compact object is $\Delta \varphi = \frac{5\pi G M} {c^2 r_c}$, then the compact object could be identified as a conformally symmetric punctual magnetic mass.

\section*{Acknowledgements}
This work was partially supported by DGAPA-UNAM, Grant No. 113514, and Conacyt, Grant No. 166391.


%

\end{document}